\documentclass[%
reprint,
superscriptaddress,
showpacs,
preprintnumbers,
amsmath,amssymb,
aps,
prb,
longbibliography,
]{revtex4-1}
\usepackage[colorinlistoftodos]{todonotes}
\usepackage{graphicx}
\usepackage{color}
\usepackage[caption=false]{subfig}
\usepackage{dcolumn}
\usepackage{bm}
\usepackage[export]{adjustbox}
\usepackage{subfig}
\usepackage[mathlines]{lineno}
\usepackage{lineno}
\usepackage{natbib}
\newcommand{\kyle}[1]{\textcolor{black}{#1}}

\usepackage{hyperref}
\begin{document}

\title{Metric Geometry Governs Optimal Control in Driven Stokes Flows:\\ Magnetic Driving and Beyond}

\author{Kyle I. McKee}
\email{kimckee@mit.edu}
\affiliation{Department of Mathematics, Massachusetts Institute of Technology, Cambridge, Massachusetts, USA}%

\date{\today}

\begin{abstract}
In a canonical Stokes flow geometry, the Hele–Shaw cell, we show that tunable circulations induced by Lorentz forces in a conducting fluid enable particle control. We reveal that energy-optimal control paths correspond to geodesics of an \kyle{emergent} Riemannian metric \kyle{defined over the fluid domain}, which are time-optimal under a maximum-power constraint. Subject to random boundary forcing, particle paths exhibit metric-governed anisotropic diffusion. Our geometric concepts governing optimal control, though developed explicitly for circulation-driven flows, generalize to generic driven Stokes flows and so elucidate recent observations in a three-dimensional context.
\end{abstract}

\maketitle


\emph{Introduction–} The ability to control fluid flow in micro- and nano-fluidic devices is imperative in a host of applications from biotechnology and diagnostics \cite{stone2004engineering,beebe2002physics} to biodefense \cite{whitesides2006origins,mondal2020microfluidics} and bubble-based computing \cite{prakash2007microfluidic}. One key objective of microfluidic systems is to mix fluids efficiently, a task made difficult by the fact that low Reynolds number flows remain laminar. To this end, passive \cite{stroock2002chaotic} and active\cite{bau2001minute,Aref_1984,evans1997planar,ottino1989kinematics} mixing strategies have been explored. Another key objective of microfluidic devices is the manipulation of passive entities, which is critical for applications including cell sorting, material synthesis, and diagnostic assays\cite{booth2025controllable,schneider2011algorithm,amoudruz2025contactless,kislaya2025particle,karimi2013hydrodynamic}. The degrees of freedom that may be leveraged for control usually include: (1) flow rates at each device inlet\cite{schneider2011algorithm,kislaya2025particle}, which prescribe the pressure-driven component of the flow; and (2) applied electric field strengths which govern the electro-osmotic contribution to the flow \cite{stone2004engineering}. Flows realizable via these controls are are \kyle{limited; in particular, in thin-gaps of constant thickness and material properties, these flows are necessarily circulation-free. This restriction significantly limits the range of possible flow-fields.}

\kyle{It was recently shown that magnetically-driven flows overcome this limitation and enable tuneable circulation around obstacles in thin-gap flows\cite{mckee2024magnetohydrodynamic}. As compared to flows driven by inlets and outlets, Lorentz-force-driven flows offer several advantages. First, they require only electrical stimuli and no moving components. Second, they markedly expand the class of realizable flow-fields by enabling circulation\footnote{Note that these flows can be superposed with pressure-driven flows to generate a broad new class of flows.}. Third, magnetically-driven flows can generate velocity in the device interior (away from inlets and outlets) without inducing a large global velocity field, which may be useful in micro-assembly when one seeks to manipulate one particle without disturbing others. Moreover, such flows may provide energetic advantages relative to their pressure-driven counterparts.}

The first aim of this Letter is to demonstrate that tuneable circulation \cite{mckee2024magnetohydrodynamic} around obstacles constitutes a new degree of freedom for particle control. To this end, we focus on flows in Hele-Shaw (HS) cells, an analogue system for studying flows in microfluidic devices and porous media. HS flow \cite{hele1898flow} exists between two closely-spaced parallel plates\kyle{; notably, in pressure-driven flows, the flow exhibits a parabolic profile across the gap when the product of the Reynolds number and the cell aspect ratio is small.} The depth-averaged flow corresponds to a two-dimensional potential flow. A bubble or particle moves, to a first approximation, in proportion to the local flow velocity computed in its absence\cite{pozrikidis1994motion,schneider2011algorithm,booth2025motion}; we take this constant of proportionality to be unity in our analysis, without loss of generality. A major limitation of traditional HS flows \footnote{We consider Hele-Shaw cells with homogeneous electrical properties and cell thickness; cells with heterogeneous properties can support vortical flows\cite{mirzadeh2020vortices,boyko2021microscale,ajdari1995electro,ajdari2000pumping}.}–driven by pressure or electro-osmotic forces–is that they possess zero circulation, which severely restricts the class of realizable flow streamline patterns \cite{van1982album}. \kyle{Such flows are circulation-free because they are driven by the gradient of physical single-valued functions (pressure and electrical potential).} Recent theoretical\cite{mckee2024magnetohydrodynamic} and experimental\cite{mckee2025potential} work demonstrated how Lorentz forces may be leveraged to generate tuneable circulations in HS cells filled with a conducting fluid. Herein, we demonstrate that these —or other \footnote{One interesting setting where potential flow with circulation emerges is in rapidly rotating flow satisfying quasi-geostrophic balance. Away from patches of relative vorticity, the flow is a potential flow with circulation; the circulation magnitude around each patch is prescribed by its net vorticity content \cite{pullin1992contour}.}— potential flows with circulation facilitate particle control, and that optimal control is governed by a metric induced on the fluid domain. 
 
The second aim of this Letter is to show that our geometric concepts, which govern optimal control, extend to generic Stokes flows. \kyle{Particularly, the framework developed for deducing the optimal control applies to any Stokes flow generated by other means (e.g., boundary-driven flows) if the fluid geometry is fixed in time. To illustrate this point, we show how the framework applies to a particular 3D Stokes flow and, in doing so, elucidate recently reported phenomena\cite{amoudruz2025contactless}.}

\emph{Physical and Mathematical Description–} Consider the flow of a conducting fluid through a HS cell containing a collection of conducting obstacles $B_k$ for $k\in\{1,2,\cdots,N\}$, that is fully immersed in a transverse magnetic field $\textbf{B}=B_0 \hat{\boldsymbol{z}}$ \cite{mckee2024magnetohydrodynamic}. By introducing electrical current between obstacles in the flow, circulation is induced around each obstacle with magnitude\cite{mckee2025potential}
\begin{equation} \Gamma_k=-\frac{I_k B_0 h}{12\mu},\label{eq:circulaitonfo}
\end{equation}
where $I_k$ is the electrical current leaving the $k^{\mathrm{th}}$ obstacle, $h$ is the cell thickness, and $\mu$ the viscosity of the conducting fluid. \kyle{Note that we consider depth-averaged velocity in the low Hartmann number limit.} By adjusting the current exiting each body, one can tune the corresponding circulation, as was demonstrated in recent experiments\cite{mckee2025potential}. Current conservation imposes the constraint $\sum_{k=1}^{N}I_k=0$; the circulation can thus be freely prescribed around  $N-1$ obstacles, whereas $\Gamma_N$ is constrained such that $\Gamma_N=-\sum_{k=1}^{N-1}\Gamma_k$.

The flow velocity can be represented as the gradient of a harmonic potential, $\boldsymbol{u}\equiv(u_x,u_y)^T=\nabla \phi$, which is the real part of a complex analytic function of the variable $z=x+\mathrm{i}y$, denoted $W(z)=\phi +\mathrm{i}\psi$, in the fluid domain defined by $D\equiv \mathbb{C}\backslash \cup_{k=1}^N B_k$. This analytic function has the following series expansion\cite{mckee2024magnetohydrodynamic},
 \begin{equation}\label{eq:laurent}
W(z;\boldsymbol{\Gamma})=\sum_{k=1}^{N-1}\frac{\mathrm{i}\Gamma_k}{2\pi}\log{\left(\frac{z-z_k}{z-z_N}\right)}+\sum_{k=1}^{N}\sum_{j=1}^{\infty}\frac{a^{(k)}_j}{\left(z-z_k\right)^j},
\end{equation}
where $\boldsymbol{\Gamma}=\left(\Gamma_1, \Gamma_2, \cdots, \Gamma_{N-1}\right)^T$ and the complex velocity is given by $u\equiv u_x+\mathrm{i}u_y=\overline{dW/dz}$. The series comprises an \emph{a priori} known multi-valued part (logarithm terms) supplemented by a Laurent series centered within each body at points $z_k\in \mathrm{int}\left(B_k\right)$, with coefficients $\{a_j^{\left(k\right)}\}$ determined by enforcing impermeability at each obstacle surface; mathematical details for finding $\{a_j^{\left(k\right)}\}$ are presented in prior work\cite{mckee2024magnetohydrodynamic}\footnote{In practice the upper limit of the $j$-summation is truncated at $N_\mathrm{L}$ terms. In canonical geometries, the solution can be written exactly in terms of the Schottky-Klein prime function\cite{crowdy2020solving}.}. Although these circulatory flows can be superposed with pressure-driven flows\footnote{for example, a free stream $U\in \mathbb{C}$ can be incorporated by augmenting $W(z)$ in \eqref{eq:laurent} with the term $\overline{U}z$.}, we restrict our attention to circulation-driven flows. 

For reasons that will become clear as we proceed, we note that \eqref{eq:laurent} can be expressed as a linear combination of $N-1$ basis potential functions, $W_k(z)$, each describing a flow with $\Gamma_j=\delta_{jk}-\delta_{jN}$ for $k<N$, so that 
\begin{equation} \label{eq:basispo} W(z;\boldsymbol{\Gamma})=\sum_{k=1}^{N-1}\Gamma_k W_k(z).
\end{equation}.

\emph{Formulation of Control Problem–} Suppose we wish to control a particle, so that it follows a prescribed trajectory $\boldsymbol{x}(t)=\left(x(t),y(t)\right)^T$, by an appropriate choice of $\boldsymbol{\Gamma}(t)$ for $t\in[0,1]$. Mathematically, we require
\begin{equation}\label{eq:controleq}
    \boldsymbol{u}(\boldsymbol{x}(t))=\dot{\boldsymbol{x}}(t),
\end{equation}
which, in terms of the basis functions \eqref{eq:basispo}, becomes

\begin{equation}\label{eq:fulleqgammelec}
\Omega(\boldsymbol{x}) \boldsymbol{\Gamma}(t)=\dot{\boldsymbol{x}}(t),
\end{equation}
where $\Omega_{1k}=\mathrm{Re}\left\{\overline{W_k'\left(z\right)}\right\}$ and $\Omega_{2k}=\mathrm{Im}\left\{\overline{W_k'(z)}\right\}$, for $k\in\{1,2\cdots,N-1\}$, and $z=x(t)+\mathrm{i}y(t)$. For the trajectory $\boldsymbol{x}(t)$ to be controllable, a solution to \eqref{eq:fulleqgammelec} must exist for all times. We now examine when such a solution exists. 
A unique solution exists if $\Omega$ is an invertible matrix for all times, which is only even possible if $N=3$; in such a case, the solution is given by $\boldsymbol{\Gamma}=\Omega^{-1}\dot{\boldsymbol{x}}$. For $N>3$ there is not a unique solution; instead, either \emph{no solutions} or \emph{infinitely many solutions} exist. \kyle{While we will not focus on this case, it is worth noting that multiple particles can be controlled within the same framework by concatenating matrices. For example, to control trajectories $\boldsymbol{x}_1(t)$ and $\boldsymbol{x}_2(t)$ simultaneously, \eqref{eq:fulleqgammelec} becomes $[\Omega(\boldsymbol{x}_1);\Omega(\boldsymbol{x}_2)]\boldsymbol{\Gamma}=[\dot{\boldsymbol{x}}_1(t);\dot{\boldsymbol{x}}_2(t)]$\footnote{The semicolon signals vertical stacking.}.}

\begin{figure}
\centering
\includegraphics[width=0.99\linewidth]{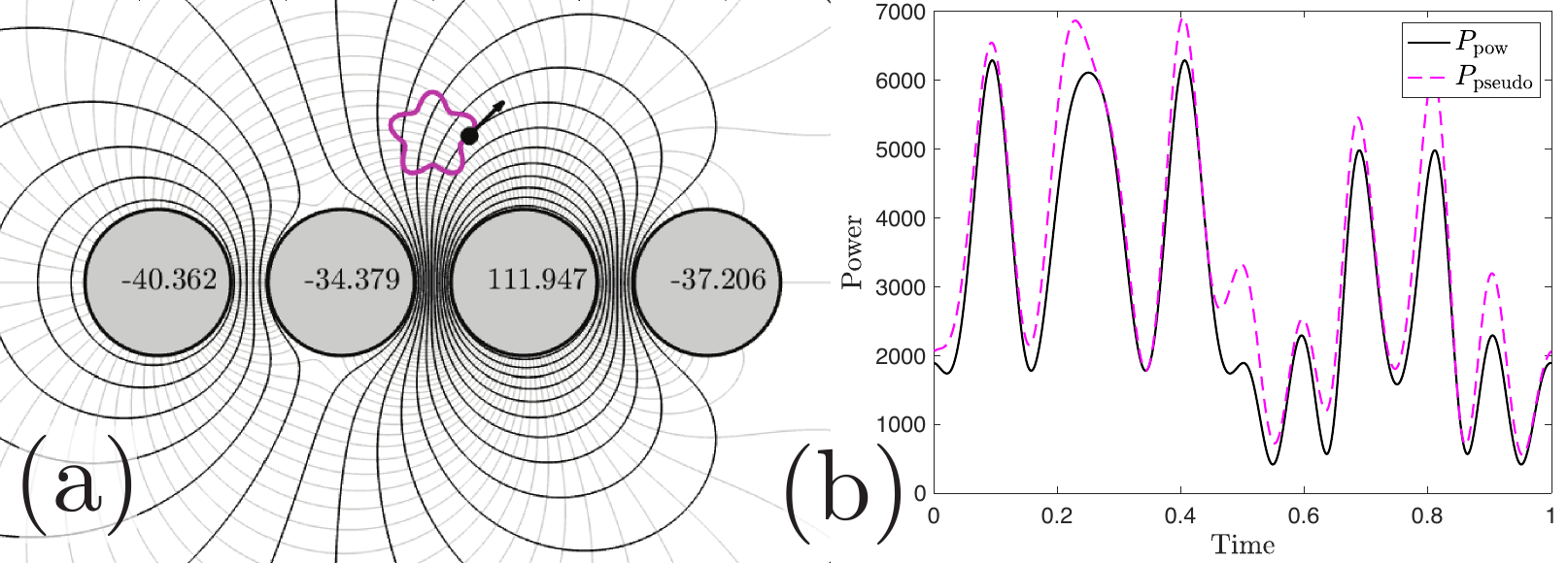}
\caption{Control of an \emph{a priori} specified trajectory. (a) A star-shaped trajectory (magenta) is controlled by varying the circulation around four conductors. Instantaneous streamlines (black), potential contours (grey), and power-optimal circulations are displayed. (b) Electrical power of pseudo-inverse and the power-optimal controls. }\label{fig:prior}
\end{figure}
\begin{figure*}
\centering
\includegraphics[width=0.95\linewidth]{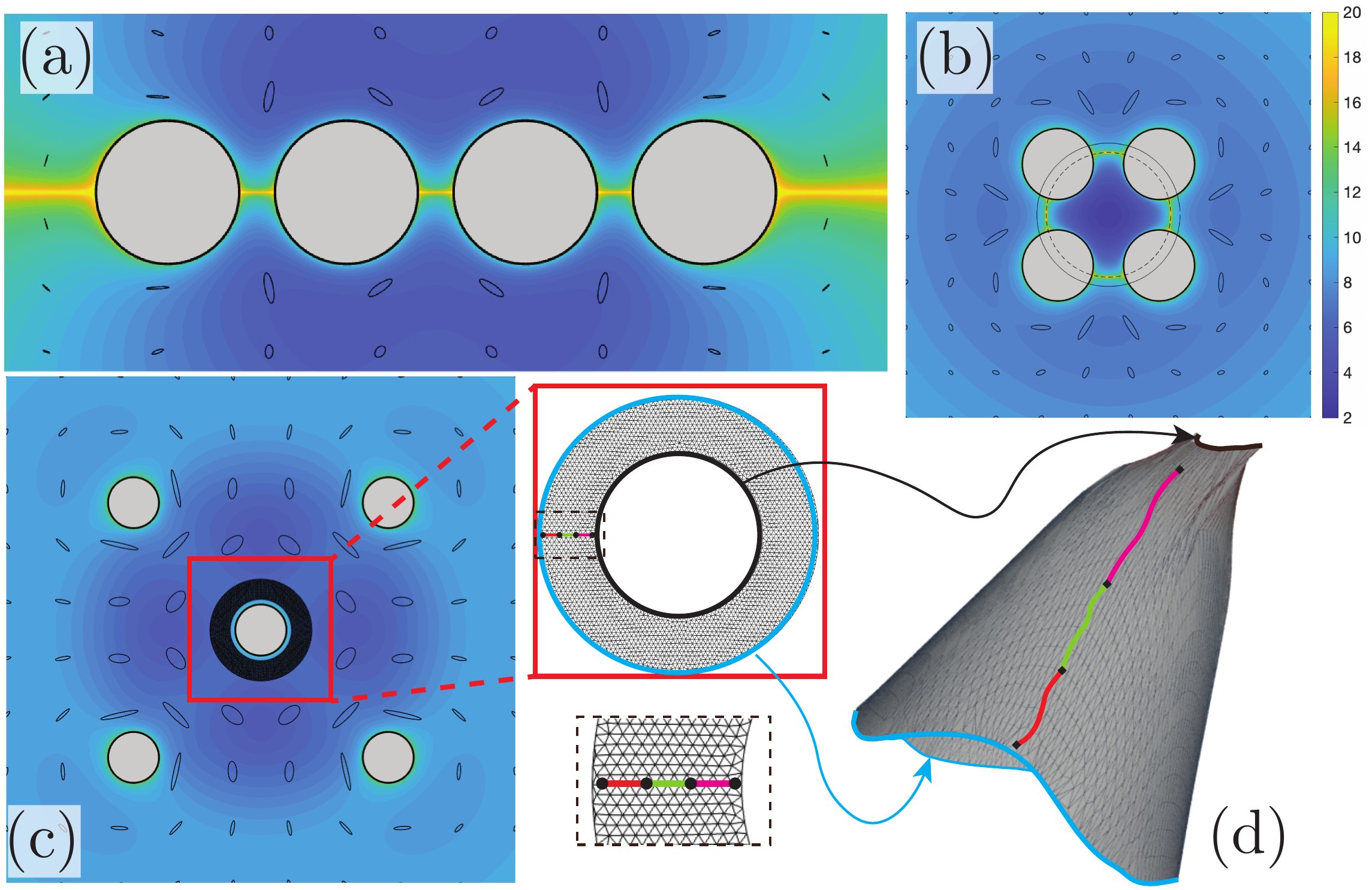}
\caption{Metric visualization. (a),(b),(c) For three conductor geometries, $\log{|\sigma_{\mathrm{max}}\left(g\right)|}$ is depicted in colour, revealing how expensive control is across the domain. Each overlaid ellipse shows the locus of points a particle may be displaced to, from its center, using a fixed amount of energy, as is determined by the local metric 
$g$. In (b) circular conductors of radius $r=0.7$ are centered on a circle of radius $R=\sqrt{2}$ (solid). A dashed circle of radius $\sqrt{R^2-r^2}$ corresponds to the singularity of $\sigma_{\mathrm{max}}\left(g\right)$ predicted by $\eqref{eq:mobius}$. (c) The addition of a central circle breaks the conformal equivalence to a geometry of the type depicted in (a), thus eliminating singularities of $\sigma_{\mathrm{max}}$. (d) Approximate isometric immersion in $\mathbb{R}^3$ of the metric $g$ over an annular region from (c). As the inner annular radius is approached, lengths (red, green, magenta) become dramatically stretched in the normal direction, such that the immersion resembles a smoothed funnel.
}\label{fig:metricvisual}
\end{figure*}

\emph{Optimal Control when $\Omega$ has Full Rank–} Away from special lines of symmetry (to be discussed later), the matrix $\Omega$ will generally possess full-rank at each point in the domain. Assuming that $\Omega$ has full rank and $N>3$, there are infinitely many solutions to \eqref{eq:fulleqgammelec}. A natural way to single out one solution, in such underdetermined systems, is to select that which minimizes the Euclidean-norm via the pseudo-inverse, $\boldsymbol{\Gamma}_{\mathrm{pseudo}}=\Omega^T\left(\Omega \Omega^T\right)^{-1} \dot{\boldsymbol{x}}(t)$. While the Euclidean-norm minimizer qualitatively possesses a small power, we shall do better by constructing a control that actually minimizes the instantaneous electrical power expenditure, as follows. Minimizing electrical power expenditure is useful for energy savings and to avoid excessive heating. By direct manipulation (see Appendix \ref{Powerderiv}), it can be shown that the electrical power required to realize a given circulation vector is given by,
\begin{equation}\label{eq:powerexpr}
P=-\alpha\sum_{k=1}^{N-1}\sum_{j=1}^{N-1} \Gamma_k\Gamma_j \mathrm{Im}\left\{ W_j(\partial B_k)\right\}\equiv\alpha\boldsymbol{\Gamma}^T M \boldsymbol{\Gamma},
\end{equation}
where $\alpha=\frac{144\mu^2}{h^2\sigma B_0^2}$ and $M$ is symmetric and positive definite\footnote{Symmetry of $M$ follows from the reciprocity of the Laplace equation, stated as follows, $\int_{\partial B_j}\psi_k(\partial B_j)\frac{\partial \psi_j}{\partial n}ds=\int_{\partial B_k}\psi_j(\partial B_k)\frac{\partial \psi_k}{\partial n}ds$. Noting that the circulation is unity $\int_{\partial B_k}\frac{\partial \psi_k}{\partial n}ds=1$, one finds that $\psi_j(\partial B_k)=\psi_k(\partial B_j)$ or $M_{jk}=M_{kj}$. Positive definiteness follows from Appendix \ref{Powerderiv}.}. The control that minimizes the power expenditure satisfies (\ref{eq:fulleqgammelec}) while simultaneously minimizing (\ref{eq:powerexpr}). Using a Lagrange multiplier approach, we derive the power-minimizing control (see Appendix \ref{LagrangeDeriv}) as follows,
\begin{equation}\label{eq:powermincontrolfun}
    \boldsymbol{\Gamma}(t)=M^{-1}\Omega^T\left(\Omega M^{-1} \Omega^T\right)^{-1}\dot{\boldsymbol{x}}(t).
\end{equation}
If $M$ is a multiple of the identity, \eqref{eq:powermincontrolfun} reduces to $\boldsymbol{\Gamma}_{\mathrm{pseudo}}$.

We showcase \eqref{eq:powermincontrolfun} by controlling the star-shaped trajectory depicted in Fig.\ref{fig:prior}(a). In panel (b), the power-minimizing control (\ref{eq:powermincontrolfun}) outperforms the pseudo-inverse solution labeled $\boldsymbol{\Gamma}_{\mathrm{pseudo}}$ at each instant. By construction, (\ref{eq:powermincontrolfun}) outperforms \emph{any} other control. We now investigate how the geometry of conductors governs controllability.

\emph{Assessing Controllability–}
Utilizing the control (\ref{eq:powermincontrolfun}) at each instant, the power expenditure \eqref{eq:powerexpr} simplifies to $P=\dot{\boldsymbol{x}}^Tg\dot{\boldsymbol{x}}$, where $g=\alpha \left(\Omega M^{-1} \Omega^T\right)^{-1}$ is positive definite and symmetric. Controllability is then characterized by the spectral norm of $g$,  $\sigma_{\mathrm{max}}\left(g\right)=\mathrm{max}_{||\dot{\boldsymbol{x}}||=1} \left(\dot{\boldsymbol{x}}^Tg\dot{\boldsymbol{x}}\right)$, the maximum power (over all angles) required to induce a unit velocity at each point. A plot of $\sigma_{\mathrm{max}}\left(g\right)$ in the fluid domain serves as a map of controllability: where it becomes large, there is at least one direction along which it is very difficult to induce flow. Such plots are included in colour in Fig. \ref{fig:metricvisual} and discussed as follows. 

\emph{Uncontrollable Trajectories: Rank-Deficiency and Symmetry–} If $\Omega$ becomes rank-deficient at a point, as can result from symmetry, then (\ref{eq:controleq}) only has a solution if the target velocity lives in the column-space of $\Omega$. As such points are approached, the power required to induce fluid velocity in a direction orthogonal to the columns of $\Omega$ diverges, as is reflected in plots of $\sigma_{\mathrm{max}}\left(g\right)$. 

Letting $\alpha=1$ for simplicity, $\sigma_{\mathrm{max}}\left(g\right)$ is depicted in colour for three distinct conductor configurations, in panels (a)-(c) of Fig. \ref{fig:metricvisual}. In the four-circle configuration of panel (a), $\sigma_{\mathrm{max}}\left(g\right)$ diverges on the real axis. As follows from the symmetry of this particular configuration, fluid velocity cannot be induced \emph{along} the real axis using circulation. Thus, the columns of $\Omega$ become degenerate and $\Omega$ becomes rank-deficient there \footnote{$\mathrm{Re}\left\{\overline{W_k'(z)}\right\}=0$ for $z\in \mathbb{R}$ and all $k\in \{1,2,\cdots, N-1\}$}. Only trajectories $\boldsymbol{x}(t)$ that cross the real axis normally are permissible. Hence, a generic trajectory connecting a point in the lower half plane to one in the upper half plane is not controllable, unless it happens to cross the real axis normally. 

The configuration in Fig. \ref{fig:metricvisual} (b) comprises four circles of radius $r=0.7$ centered on a circle of radius $R=\sqrt{2}$. We observe that $\sigma_{\mathrm{max}}\left(g\right)$ diverges along a circle of radius $R_c<R$. We rationalize this divergence by first noting that the geometry is related via the Möbius map,
\begin{equation}\label{eq:mobius}
    m(z)=\mathrm{i}\frac{z+\sqrt{R^2-r^2}}{z-\sqrt{R^2-r^2}},
\end{equation}
to a set of circles with their centers on the real axis. By the conformal invariance of the Laplace equation, we consider the flow problem in the mapped geometry and revert to the physical plane by inverting $m(z)$ thereafter. Hence, the pre-image of the real axis under $m(z)$ is a line tangent to which no fluid flow can be induced. Möbius maps take circles to circles (or straight lines) and the pre-image of the real axis is a circle of radius $R_c=\sqrt{R^2-r^2}$. A dashed circle of radius $\sqrt{1.51}$ is plotted in Fig. \ref{fig:metricvisual}(b), which coincides precisely with the divergence of $\sigma_{\mathrm{max}}\left(g\right)$. Generally, \emph{any} geometry related by a conformal mapping to a configuration of circles (possibly of unequal radii) centered on the real axis contains a line where $\sigma_{\mathrm{max}}\left(g\right)$ diverges, at the pre-image of the real axis under the map\footnote{In any geometry wherein all obstacles intersect the real axis, while possessing reflectional symmetry about this axis (i.e.,  $D=\overline{D}$), $\sigma_{\mathrm{max}}$ is necessarily singular along this line. This phenomena is thus by no means restricted to the case of circular conductors.}.  

\begin{figure*}
\centering
\includegraphics[width=0.9\linewidth]{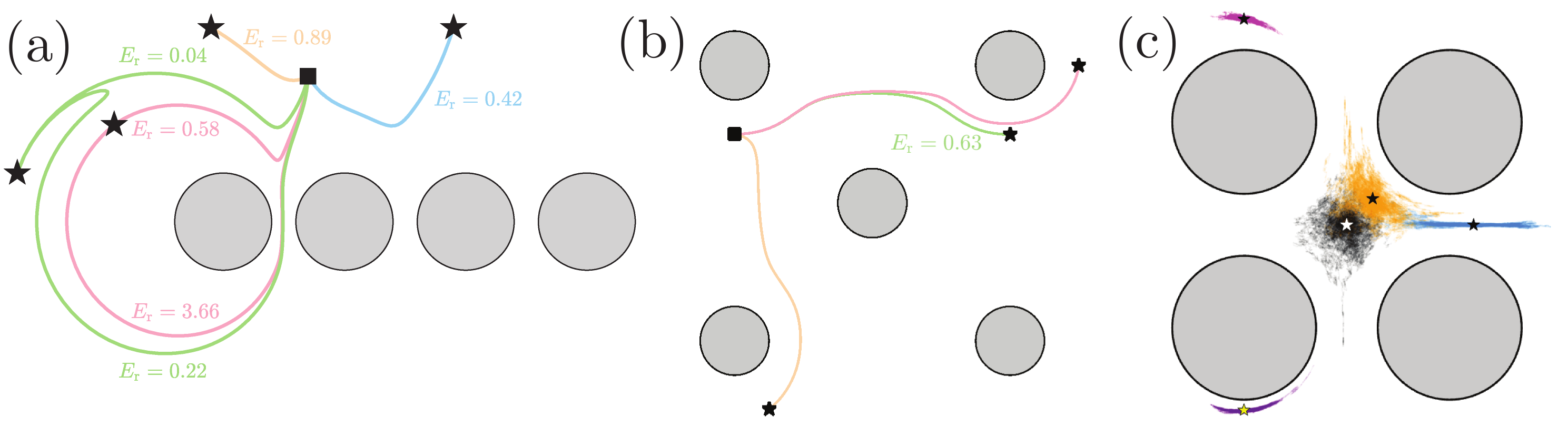}
\caption{ (A) Four circular conductors are centered on the real axis.
Geodesics between \kyle{$\boldsymbol{x}_A$ (square) and four target points $\boldsymbol{x}_B$ (stars) are depicted} and labelled with $E_r$, the ratio of geodesic energy to that of a straight-line trajectory. Two green and two red trajectories connect the same endpoints. (B) Geodesics around five circular conductors, centered on the corners and center of a square. (C) Particles are released from five initial points (stars) in the presence of random circulation forcing. Each colour corresponds to 100 overlaid equal-time trajectories starting from the corresponding star.
}\label{fig:geodesics}
\end{figure*}

\emph{Optimal Control–} We have considered the control of \emph{a priori} specified trajectories, and found the power-minimizing control (\ref{eq:powermincontrolfun}) valid if: (1) $\Omega$ retains full-rank along the trajectory \emph{or} (2) where $\Omega$ loses full rank, $\dot{\boldsymbol{x}}\left(t\right)$ remains in the columnspace of $\Omega$. If both criteria fail, the trajectory is not controllable. While $\sigma_{\mathrm{max}}\left( g\right)$ reveals controllability in the domain, so trajectories may be designed to avoid costly regions, one is left wondering if there exists an energy-optimal trajectory between points.

\emph{Given an initial point, $\boldsymbol{x}_A$, and a target point to be reached at $t=1$, $\boldsymbol{x}_B$, what trajectory $\boldsymbol{x}\left(t\right)$ and control $\boldsymbol{\Gamma}\left(t\right)$ minimize the net energy expenditure?} Utilizing (\ref{eq:powermincontrolfun}) at each instant, the total energy required to control a generic trajectory is given by the following functional,
\begin{eqnarray}
E[\boldsymbol{x}\left(t\right)]=\int_{0}^{1}\boldsymbol{\dot{x}}(\tau)^T g(\boldsymbol{x}(\tau))\boldsymbol{\dot{x}}(\tau) d\tau.\label{eq:functional}\label{eq:energyeq}
\end{eqnarray}
Subject to the endpoint constraints $\boldsymbol{x}(0)=\boldsymbol{x}_A$ and $\boldsymbol{x}(1)=\boldsymbol{x}_B$, the energy is extremized by a trajectory satisfying the Euler-Lagrange equations corresponding to the action \eqref{eq:energyeq}. The Euler-Lagrange equations describing $\boldsymbol{x}(t)$ reduce as follows (see Appendix \ref{geodesicderivations}),
\begin{eqnarray}
\ddot{x}_k+\Gamma^{k}_{rs}\dot{x}_r\dot{x}_s=0,\label{eq:geodesic}\\
\Gamma^{k}_{rs} \equiv \frac{1}{2}g^{-1}_{kq}\left(\frac{\partial g_{qr}}{\partial x_s}+\frac{\partial g_{qs}}{\partial x_r}-\frac{\partial g_{ r  s}}{\partial x_q}\right)\label{eq:christoffel},
\end{eqnarray}
where repeated indices indicate summation and $r,s\in \{1,2\}$. Trajectories satisfying (\ref{eq:geodesic}) are geodesics corresponding to the two-dimensional Riemannian metric $g$. It is worth noting that our solution indeed satisfies the Pontryagin optimality conditions \cite[p. 19]{pontryagin2018mathematical}. 

To recap, solutions to a sequence of $N-1$ Laplace problems over $D$ served to define a two-dimensional Riemannian metric, $g$. Shortest paths measured with respect to this metric define energy-minimizing trajectories between points. It is worth mentioning a simple parallel with the standard program in general relativity wherein a metric is first computed as the solution of a PDE (albeit Einstein's equations\cite{carroll2019spacetime} rather than Laplace's equation) and subsequently particle trajectories follow geodesics atop this metric. Of course, there are many caveats, one being that time does not enter our metric. 

\emph{Metric Structure–} Generic features of $g$ follow from the boundary conditions satisfied by the fluid flow. Impermeability at each obstacle boundary yields infinite stretching of length, as measured with respect to $g$, in the direction \emph{normal} to the boundary. In particular, lengths scale as $\log{|n|^{-1}}$, where $n$ is the normal distance to the boundary \footnote{Consider a decomposition of the metric, near a point on the boundary of a conductor, into tangent and normal directions to its surface. Let $n$ denote the normal distance from the boundary and $t$ the tangent distance along it. Since $g^{-1}=\Omega M^{-1} \Omega^T$, impermeability implies that locally the normal component of the metric scales as $g^{-1}_{nn}\sim n^2/C+ \it{o}\left(n^2\right)$ for some constant $C$. Meanwhile, there is no metric singularity in the tangent direction, $g^{-1}_{tt}\sim C'+ \it{o}\left(1\right)$. Hence, $g_{nn}\sim C/n^2$. Hence the length measured orthogonal to the boundary between $n=\epsilon$ to $n=x$ becomes $\int_{\epsilon}^{x}\sqrt{g_{nn}}dn=\int_{\epsilon}^{x}\left(1/n\right)dn=\log{\left(x/\epsilon\right)}$. Normal distances thus blowup logarithmically as the boundary is approached.
}. There is no such singularity in the \emph{tangent} direction. The metric can be visualized in the plane by plotting ellipses traced out by $\boldsymbol{q}$ satisfying $\boldsymbol{q}^Tg(\boldsymbol{x})\boldsymbol{q}=l^2$ for $l^2>0$ fixed, which depict the locus of points reachable by a particle starting at its center with a fixed energy $l^2$. Ellipse semiaxes are given by $r_{1,2}=\sqrt{l^2/\lambda_{1,2}}$, where $\lambda_{1,2}$ are the eigenvalues of $g$. Such ellipses overlay the plots in panels (a),(b) and (c) of Fig. \ref{fig:metricvisual}. At conducting boundaries, ellipses degenerate to slits tangent to the boundary, as it becomes infinitely costly to move normally to these surfaces. In central regions, ellipses approach circularity. Far from conducting boundaries, ellipses decay in size since velocity fields decay at infinity. It is thus costly to control particles far away from the conducting obstacles. 

\emph{Isometric Immersion–}  
A global isometric immersion of our metric as a surface in $\mathbb{R}^K$ is guaranteed for $K$ sufficiently large \cite{nash1954c}, though it may be crumpled and require $K>3$. However, because our metric is analytic, there exists\cite[p.4]{han2006isometric} a \emph{local} isometric immersion in $\mathbb{R}^3$. In Fig. \ref{fig:metricvisual}(d), we visualize an approximately isometric immersion of the annular region depicted in (c). The extrinsic geometry of the immersion is selected via a curvature-penalizing scheme developed by Chern et al. \cite{chern2018shape}. If a global isometric immersion exists, we expect it to resemble a set of volcanoes–one per conductor–of infinite height.

\emph{Anisotropic Diffusion–}
The metric can also be visualized by randomly forcing particles in the flow as in Fig. \ref{fig:geodesics}(c). Let $\boldsymbol{\zeta}=M^{\frac{1}{2}}\boldsymbol{\Gamma}$ be mean-zero Gaussian noise\footnote{Sampling $M^{\frac{1}{2}}\boldsymbol{\Gamma}$ rather than $\boldsymbol{\Gamma}$ from a mean-zero Gaussian distribution ensures that equal-power ($P=\boldsymbol{\Gamma}^T M \boldsymbol{\Gamma}$) configurations are explored equally. When the distribution has variance $\sigma^2$, the electrical power follows a Chi-squared distribution with $\mathbb{E}(P)=\left(N-1\right)\sigma^2$.}–which may be realized by applying noisy voltages to the conductors–sampled on time intervals short enough that convective motions during each interval are small compared to obstacle radii. Since $\dot{\boldsymbol{x}}=\Omega M^{-\frac{1}{2}}\boldsymbol{\zeta}$, particle dynamics are diffusive, corresponding to an anisotropic diffusion tensor $g^{-1}=\Omega M^{-1}\Omega^T$. Overlaying 100 trajectories from various starting points (stars) in Fig. \ref{fig:geodesics}(c) reveals the metric structure. Near the origin (black), particles spread isotropically, while 
elsewhere (cyan) the spread is primarily in the direction of the minimal eigenvalue direction of $g$, mimicking the ellipses plots of Fig. \ref{fig:metricvisual}. Various metrics can be visualized by changing the geometry of conductors.

\emph{Geodesics–} The features of the metric discussed above are reflected in the geodesics plotted for two geometries in Fig. \ref{fig:geodesics}, which were attained numerically via a shooting method \footnote{Our numerical method utilized the Hamiltonian form of the Euler-Lagrange equations \eqref{eq:geodesic} wherein $H=\left(\boldsymbol{p}^Tg^{-1}\boldsymbol{p}\right)/4$ and $\boldsymbol{p}=2g\dot{\boldsymbol{x}}$. In this form, we verify convergence of our integrations by monitoring the Hamiltonian conservation along trajectories; for geodesics included in the paper, Hamiltonian fluctuations are less that 1\% along trajectories.}. 
In each case, the initial point is indicated by a square and target points by stars. $E_r$ denotes the ratio of the energy consumed by a geodesic to that of an equal-time straight-line trajectory. When $E_r$ can be defined, geodesics significantly outperform straight-line trajectories such that $E_r<1$. In panel (a), one geodesic requires just $4\%$ of the energy of the straight-line trajectory. This remarkable savings comes from the fact that in regions where the columns of $\Omega$ are nearly parallel, inducing power in a direction orthogonal to these columns becomes extremely costly. Whereas straight-line trajectories inevitably have some velocity component in this costly direction, geodesics naturally avoid this via \eqref{eq:geodesic}. To further this point, notice that both geodesics which cross the real axis in panel (a) do so smoothly in the imaginary direction to avoid horizontal velocity contributions where we found $\Omega$ becomes rank-deficient. \kyle{As compared to typical optimization schemes, finding geodesics here is computationally simple and only requires solving an ODE. Another advantage is that geodesics are guaranteed to extremize energy whereas in optimization schemes, one needs to be careful to ensure convergence.}

\emph{Nonuniqueness}– Because $D$ is multiply-connected, there is not a \emph{unique} geodesic trajectory connecting two points. Multiple geodesics exist between two points, as is demonstrated in the green and red curves in Fig.\ref{fig:geodesics}(a). Imagine connecting the initial and target point by a rubber band that can be deformed in the fluid domain but cannot be cut to cross obstacles; geodesics extremize \eqref{eq:energyeq} among trajectories related by such deformations. Hence, there exists a local minimum for each equivalence class of curves. The geodesic with the least number of windings through the conductors tends to possess the lowest energy. Since some geodesics correspond to local rather than global energy-minimizers, some require more energy than a straightline path, as is the case in the lower red curve which has $E_r=3.66$. The globally optimal trajectory ($E_r=0.58$) belongs to a different equivalence class.

\emph{Time-Optimality–} Having found the most energy-efficient path between points, one is naturally left wondering about time optimality: \emph{which trajectory minimizes the transport time between two points, subject to a power constraint of the form $P(\tau) \leq P_{\mathrm{max}}$ for all time?} As is proved in Appendix \ref{timeoptimality}, the time-optimal trajectory corresponds to a geodesic connecting the two points traversed with $P(t)= P_{\mathrm{max}}$. Geodesics are constant-power trajectories since the integrand in \eqref{eq:geodesic} is time-independent; it is thus always possible to find such a geodesic.

\emph{Extensions of Optimal Control Framework in 3D–} We have demonstrated that tuneable circulation can be harnessed for control. Controllability is governed by a Riemannian metric that is emergent from the geometry of conducting obstacles and strongly influenced by system symmetries. Energy- and time-optimal control paths correspond to geodesics of this metric. Subject to random circulation forcing, particles diffuse anisotropically according to the metric. While we focused on potential flows with circulation, our geometric ideas apply generally to Stokes flows. \kyle{Indeed, none of the story changes when one considers a generic boundary-driven or body-force-driven Stokes flow, even in 3D, as long as the fluid domain remains fixed in time. In such Stokes flows, one can define a metric on the fluid domain which governs both optimal control and anisotropic diffusion in much the same manner as was demonstrated in our magnetically-driven flows. To illustrate, we now show how our framework may be applied in a particular 3D system.} In recent 3D experiments, particles were controlled by rotating disks mounted on $N_d$ walls of a cube \cite{amoudruz2025contactless}; subject to random disk rotations, particles were observed to diffuse on a 2D manifold. This control problem can be formulated in a manner analogous to the present Letter– with $M\in \mathbb{R}^{N_d\times N_d}$ and $\Omega\in \mathbb{R}^{3\times N_d}$ (see Appendix \ref{stokesrotat})– within which \eqref{eq:powermincontrolfun} and \eqref{eq:geodesic} hold with $g=\left(\Omega M^{-1} \Omega^T\right)^{-1}$. \kyle{Again, controllability is related to the rank of $\Omega$; since each disk can be rotated independently, a minimum of $N_d=3$ disks are needed to control a generic trajectory\footnote{In the 2D magnetic system, $N=3$ conductors really only corresponds to two degrees of freedom because of the current conservation constraint.}.} Random boundary forcing yields anisotropic diffusion according to $g^{-1}$. A plot of ellipsoids in the cube, with semiaxes corresponding to eigenvalues and eigendirections of $g^{-1}$, presumably yields flattened ellipsoids tangent to the 2D manifold observed in Fig. 3 of Amoudruz et al.\cite{amoudruz2025contactless}. \kyle{While Amoudruz et al. used a numerical optimization framework to find optimal trajectories subject to a maximum rotation rate constraint on the disks, our paths were derived subject to a maximum power (dissipation) constraint. It would be interesting in future work to compare the resulting trajectories.}

\emph{Discussion and Conclusion--} 
We have shown that in magnetically driven flows, geodesics of an emergent Riemannian metric represent optimal transport paths. Moreover, transporting a particle along these paths requires a small percentage of the energy associated with a straight-line trajectory. The same metric governs the diffusion of particles when subjected to random boundary forcing. \kyle{Flows with tunable circulations should enable enhanced mixing via chaotic advection, in a manner analogous to Aref’s blinking vortices \cite{Aref_1984}, which would be an interesting direction for future work. Moreover, potential energetic benefits of magnetically-driven flows shall be investigated elsewhere.} The framework laid out for constructing the metric and corresponding optimal geodesics extends beyond our particular example of magnetically-driven flows to generic boundary-driven or body-force-driven Stokes flows in 2D or 3D. In future studies, understanding metrics and geodesics in other Stokes flows (e.g., those driven by electromagnets \cite{qin2022confinement,uchytil2025data,karnakov2025optimal}) should elucidate the interplay between geometry and controllability and so inform system design.

\textbf{Acknowledgements.}
K.M. thanks Larry Guth and Albert Chern for useful discussions regarding metric visualization. He thanks Gwynn Elfring for illuminating discussions at an early stage of this study. He thanks Daniel Booth for sharing his knowledge of bubble motion in Hele-Shaw cells. He also thanks Jean-Christophe Nave, William Holman-Bissegger, Vakhtang Putkaradze, and Glenn Flierl for illuminating discussions.

\bibliography{main.bib}

\begin{thebibliography}{49}%
\makeatletter
\providecommand \@ifxundefined [1]{%
 \@ifx{#1\undefined}
}%
\providecommand \@ifnum [1]{%
 \ifnum #1\expandafter \@firstoftwo
 \else \expandafter \@secondoftwo
 \fi
}%
\providecommand \@ifx [1]{%
 \ifx #1\expandafter \@firstoftwo
 \else \expandafter \@secondoftwo
 \fi
}%
\providecommand \natexlab [1]{#1}%
\providecommand \enquote  [1]{``#1''}%
\providecommand \bibnamefont  [1]{#1}%
\providecommand \bibfnamefont [1]{#1}%
\providecommand \citenamefont [1]{#1}%
\providecommand \href@noop [0]{\@secondoftwo}%
\providecommand \href [0]{\begingroup \@sanitize@url \@href}%
\providecommand \@href[1]{\@@startlink{#1}\@@href}%
\providecommand \@@href[1]{\endgroup#1\@@endlink}%
\providecommand \@sanitize@url [0]{\catcode `\\12\catcode `\$12\catcode `\&12\catcode `\#12\catcode `\^12\catcode `\_12\catcode `\%12\relax}%
\providecommand \@@startlink[1]{}%
\providecommand \@@endlink[0]{}%
\providecommand \url  [0]{\begingroup\@sanitize@url \@url }%
\providecommand \@url [1]{\endgroup\@href {#1}{\urlprefix }}%
\providecommand \urlprefix  [0]{URL }%
\providecommand \Eprint [0]{\href }%
\providecommand \doibase [0]{http://dx.doi.org/}%
\providecommand \selectlanguage [0]{\@gobble}%
\providecommand \bibinfo  [0]{\@secondoftwo}%
\providecommand \bibfield  [0]{\@secondoftwo}%
\providecommand \translation [1]{[#1]}%
\providecommand \BibitemOpen [0]{}%
\providecommand \bibitemStop [0]{}%
\providecommand \bibitemNoStop [0]{.\EOS\space}%
\providecommand \EOS [0]{\spacefactor3000\relax}%
\providecommand \BibitemShut  [1]{\csname bibitem#1\endcsname}%
\let\auto@bib@innerbib\@empty
\bibitem [{\citenamefont {Stone}\ \emph {et~al.}(2004)\citenamefont {Stone}, \citenamefont {Stroock},\ and\ \citenamefont {Ajdari}}]{stone2004engineering}%
  \BibitemOpen
  \bibfield  {author} {\bibinfo {author} {\bibfnamefont {Howard~A}\ \bibnamefont {Stone}}, \bibinfo {author} {\bibfnamefont {Abraham~D}\ \bibnamefont {Stroock}}, \ and\ \bibinfo {author} {\bibfnamefont {Armand}\ \bibnamefont {Ajdari}},\ }\bibfield  {title} {\enquote {\bibinfo {title} {Engineering flows in small devices: microfluidics toward a lab-on-a-chip},}\ }\href@noop {} {\bibfield  {journal} {\bibinfo  {journal} {Annu. Rev. Fluid Mech.}\ }\textbf {\bibinfo {volume} {36}},\ \bibinfo {pages} {381--411} (\bibinfo {year} {2004})}\BibitemShut {NoStop}%
\bibitem [{\citenamefont {Beebe}\ \emph {et~al.}(2002)\citenamefont {Beebe}, \citenamefont {Mensing},\ and\ \citenamefont {Walker}}]{beebe2002physics}%
  \BibitemOpen
  \bibfield  {author} {\bibinfo {author} {\bibfnamefont {David~J}\ \bibnamefont {Beebe}}, \bibinfo {author} {\bibfnamefont {Glennys~A}\ \bibnamefont {Mensing}}, \ and\ \bibinfo {author} {\bibfnamefont {Glenn~M}\ \bibnamefont {Walker}},\ }\bibfield  {title} {\enquote {\bibinfo {title} {Physics and applications of microfluidics in biology},}\ }\href@noop {} {\bibfield  {journal} {\bibinfo  {journal} {Annual review of biomedical engineering}\ }\textbf {\bibinfo {volume} {4}},\ \bibinfo {pages} {261--286} (\bibinfo {year} {2002})}\BibitemShut {NoStop}%
\bibitem [{\citenamefont {Whitesides}(2006)}]{whitesides2006origins}%
  \BibitemOpen
  \bibfield  {author} {\bibinfo {author} {\bibfnamefont {George~M}\ \bibnamefont {Whitesides}},\ }\bibfield  {title} {\enquote {\bibinfo {title} {The origins and the future of microfluidics},}\ }\href@noop {} {\bibfield  {journal} {\bibinfo  {journal} {nature}\ }\textbf {\bibinfo {volume} {442}},\ \bibinfo {pages} {368--373} (\bibinfo {year} {2006})}\BibitemShut {NoStop}%
\bibitem [{\citenamefont {Mondal}\ \emph {et~al.}(2020)\citenamefont {Mondal}, \citenamefont {Bhavanashri}, \citenamefont {Mounika}, \citenamefont {Tuteja}, \citenamefont {Tandi},\ and\ \citenamefont {Soniya}}]{mondal2020microfluidics}%
  \BibitemOpen
  \bibfield  {author} {\bibinfo {author} {\bibfnamefont {Bhairab}\ \bibnamefont {Mondal}}, \bibinfo {author} {\bibfnamefont {N}~\bibnamefont {Bhavanashri}}, \bibinfo {author} {\bibfnamefont {SP}~\bibnamefont {Mounika}}, \bibinfo {author} {\bibfnamefont {Deepika}\ \bibnamefont {Tuteja}}, \bibinfo {author} {\bibfnamefont {Kunti}\ \bibnamefont {Tandi}}, \ and\ \bibinfo {author} {\bibfnamefont {H}~\bibnamefont {Soniya}},\ }\bibfield  {title} {\enquote {\bibinfo {title} {Microfluidics application for detection of biological warfare agents},}\ }in\ \href@noop {} {\emph {\bibinfo {booktitle} {Handbook on biological warfare preparedness}}}\ (\bibinfo  {publisher} {Elsevier},\ \bibinfo {year} {2020})\ pp.\ \bibinfo {pages} {103--131}\BibitemShut {NoStop}%
\bibitem [{\citenamefont {Prakash}\ and\ \citenamefont {Gershenfeld}(2007)}]{prakash2007microfluidic}%
  \BibitemOpen
  \bibfield  {author} {\bibinfo {author} {\bibfnamefont {Manu}\ \bibnamefont {Prakash}}\ and\ \bibinfo {author} {\bibfnamefont {Neil}\ \bibnamefont {Gershenfeld}},\ }\bibfield  {title} {\enquote {\bibinfo {title} {Microfluidic bubble logic},}\ }\href@noop {} {\bibfield  {journal} {\bibinfo  {journal} {Science}\ }\textbf {\bibinfo {volume} {315}},\ \bibinfo {pages} {832--835} (\bibinfo {year} {2007})}\BibitemShut {NoStop}%
\bibitem [{\citenamefont {Stroock}\ \emph {et~al.}(2002)\citenamefont {Stroock}, \citenamefont {Dertinger}, \citenamefont {Ajdari}, \citenamefont {Mezic}, \citenamefont {Stone},\ and\ \citenamefont {Whitesides}}]{stroock2002chaotic}%
  \BibitemOpen
  \bibfield  {author} {\bibinfo {author} {\bibfnamefont {Abraham~D}\ \bibnamefont {Stroock}}, \bibinfo {author} {\bibfnamefont {Stephan~KW}\ \bibnamefont {Dertinger}}, \bibinfo {author} {\bibfnamefont {Armand}\ \bibnamefont {Ajdari}}, \bibinfo {author} {\bibfnamefont {Igor}\ \bibnamefont {Mezic}}, \bibinfo {author} {\bibfnamefont {Howard~A}\ \bibnamefont {Stone}}, \ and\ \bibinfo {author} {\bibfnamefont {George~M}\ \bibnamefont {Whitesides}},\ }\bibfield  {title} {\enquote {\bibinfo {title} {Chaotic mixer for microchannels},}\ }\href@noop {} {\bibfield  {journal} {\bibinfo  {journal} {Science}\ }\textbf {\bibinfo {volume} {295}},\ \bibinfo {pages} {647--651} (\bibinfo {year} {2002})}\BibitemShut {NoStop}%
\bibitem [{\citenamefont {Bau}\ \emph {et~al.}(2001)\citenamefont {Bau}, \citenamefont {Zhong},\ and\ \citenamefont {Yi}}]{bau2001minute}%
  \BibitemOpen
  \bibfield  {author} {\bibinfo {author} {\bibfnamefont {Haim~H}\ \bibnamefont {Bau}}, \bibinfo {author} {\bibfnamefont {Jihua}\ \bibnamefont {Zhong}}, \ and\ \bibinfo {author} {\bibfnamefont {Mingqiang}\ \bibnamefont {Yi}},\ }\bibfield  {title} {\enquote {\bibinfo {title} {A minute magneto hydro dynamic (mhd) mixer},}\ }\href@noop {} {\bibfield  {journal} {\bibinfo  {journal} {Sensors and Actuators B: Chemical}\ }\textbf {\bibinfo {volume} {79}},\ \bibinfo {pages} {207--215} (\bibinfo {year} {2001})}\BibitemShut {NoStop}%
\bibitem [{\citenamefont {Aref}(1984)}]{Aref_1984}%
  \BibitemOpen
  \bibfield  {author} {\bibinfo {author} {\bibfnamefont {Hassan}\ \bibnamefont {Aref}},\ }\bibfield  {title} {\enquote {\bibinfo {title} {Stirring by chaotic advection},}\ }\href {\doibase 10.1017/S0022112084001233} {\bibfield  {journal} {\bibinfo  {journal} {Journal of Fluid Mechanics}\ }\textbf {\bibinfo {volume} {143}},\ \bibinfo {pages} {1–21} (\bibinfo {year} {1984})}\BibitemShut {NoStop}%
\bibitem [{\citenamefont {Evans}\ \emph {et~al.}(1997)\citenamefont {Evans}, \citenamefont {Liepmann},\ and\ \citenamefont {Pisano}}]{evans1997planar}%
  \BibitemOpen
  \bibfield  {author} {\bibinfo {author} {\bibfnamefont {John}\ \bibnamefont {Evans}}, \bibinfo {author} {\bibfnamefont {Dorian}\ \bibnamefont {Liepmann}}, \ and\ \bibinfo {author} {\bibfnamefont {Albert~P}\ \bibnamefont {Pisano}},\ }\bibfield  {title} {\enquote {\bibinfo {title} {Planar laminar mixer},}\ }in\ \href@noop {} {\emph {\bibinfo {booktitle} {Proc. IEEE MEMS Workshop}}},\ Vol.~\bibinfo {volume} {10}\ (\bibinfo {year} {1997})\ pp.\ \bibinfo {pages} {96--101}\BibitemShut {NoStop}%
\bibitem [{\citenamefont {Ottino}(1989)}]{ottino1989kinematics}%
  \BibitemOpen
  \bibfield  {author} {\bibinfo {author} {\bibfnamefont {Julio~M}\ \bibnamefont {Ottino}},\ }\href@noop {} {\emph {\bibinfo {title} {The kinematics of mixing: stretching, chaos, and transport}}},\ Vol.~\bibinfo {volume} {3}\ (\bibinfo  {publisher} {Cambridge university press},\ \bibinfo {year} {1989})\BibitemShut {NoStop}%
\bibitem [{\citenamefont {Booth}\ and\ \citenamefont {Montenegro-Johnson}(2025)}]{booth2025controllable}%
  \BibitemOpen
  \bibfield  {author} {\bibinfo {author} {\bibfnamefont {Daniel~J}\ \bibnamefont {Booth}}\ and\ \bibinfo {author} {\bibfnamefont {Thomas~D}\ \bibnamefont {Montenegro-Johnson}},\ }\bibfield  {title} {\enquote {\bibinfo {title} {Controllable microfluidics through active droplets},}\ }\href@noop {} {\bibfield  {journal} {\bibinfo  {journal} {Physical Review Fluids}\ }\textbf {\bibinfo {volume} {10}},\ \bibinfo {pages} {094203} (\bibinfo {year} {2025})}\BibitemShut {NoStop}%
\bibitem [{\citenamefont {Schneider}\ \emph {et~al.}(2011)\citenamefont {Schneider}, \citenamefont {Mandre},\ and\ \citenamefont {Brenner}}]{schneider2011algorithm}%
  \BibitemOpen
  \bibfield  {author} {\bibinfo {author} {\bibfnamefont {Tobias~M}\ \bibnamefont {Schneider}}, \bibinfo {author} {\bibfnamefont {Shreyas}\ \bibnamefont {Mandre}}, \ and\ \bibinfo {author} {\bibfnamefont {Michael~P}\ \bibnamefont {Brenner}},\ }\bibfield  {title} {\enquote {\bibinfo {title} {Algorithm for a microfluidic assembly line},}\ }\href@noop {} {\bibfield  {journal} {\bibinfo  {journal} {Physical review letters}\ }\textbf {\bibinfo {volume} {106}},\ \bibinfo {pages} {094503} (\bibinfo {year} {2011})}\BibitemShut {NoStop}%
\bibitem [{\citenamefont {Amoudruz}\ \emph {et~al.}(2025)\citenamefont {Amoudruz}, \citenamefont {Karnakov},\ and\ \citenamefont {Koumoutsakos}}]{amoudruz2025contactless}%
  \BibitemOpen
  \bibfield  {author} {\bibinfo {author} {\bibfnamefont {Lucas}\ \bibnamefont {Amoudruz}}, \bibinfo {author} {\bibfnamefont {Petr}\ \bibnamefont {Karnakov}}, \ and\ \bibinfo {author} {\bibfnamefont {Petros}\ \bibnamefont {Koumoutsakos}},\ }\bibfield  {title} {\enquote {\bibinfo {title} {Contactless precision steering of particles in a fluid inside a cube with rotating walls},}\ }\href@noop {} {\bibfield  {journal} {\bibinfo  {journal} {Journal of Fluid Mechanics}\ }\textbf {\bibinfo {volume} {1014}},\ \bibinfo {pages} {A15} (\bibinfo {year} {2025})}\BibitemShut {NoStop}%
\bibitem [{\citenamefont {Kislaya}\ \emph {et~al.}(2025)\citenamefont {Kislaya}, \citenamefont {Samant}, \citenamefont {Veenstra}, \citenamefont {Tam},\ and\ \citenamefont {Westerweel}}]{kislaya2025particle}%
  \BibitemOpen
  \bibfield  {author} {\bibinfo {author} {\bibfnamefont {Ankur}\ \bibnamefont {Kislaya}}, \bibinfo {author} {\bibfnamefont {Aniket~Ashwin}\ \bibnamefont {Samant}}, \bibinfo {author} {\bibfnamefont {Peter}\ \bibnamefont {Veenstra}}, \bibinfo {author} {\bibfnamefont {Daniel~SW}\ \bibnamefont {Tam}}, \ and\ \bibinfo {author} {\bibfnamefont {Jerry}\ \bibnamefont {Westerweel}},\ }\bibfield  {title} {\enquote {\bibinfo {title} {Particle manipulation in {H}ele--{S}haw flow with programmable hydrodynamics},}\ }\href@noop {} {\bibfield  {journal} {\bibinfo  {journal} {Physics of Fluids}\ }\textbf {\bibinfo {volume} {37}} (\bibinfo {year} {2025})}\BibitemShut {NoStop}%
\bibitem [{\citenamefont {Karimi}\ \emph {et~al.}(2013)\citenamefont {Karimi}, \citenamefont {Yazdi},\ and\ \citenamefont {Ardekani}}]{karimi2013hydrodynamic}%
  \BibitemOpen
  \bibfield  {author} {\bibinfo {author} {\bibfnamefont {A}~\bibnamefont {Karimi}}, \bibinfo {author} {\bibfnamefont {S}~\bibnamefont {Yazdi}}, \ and\ \bibinfo {author} {\bibfnamefont {AM}~\bibnamefont {Ardekani}},\ }\bibfield  {title} {\enquote {\bibinfo {title} {Hydrodynamic mechanisms of cell and particle trapping in microfluidics},}\ }\href@noop {} {\bibfield  {journal} {\bibinfo  {journal} {Biomicrofluidics}\ }\textbf {\bibinfo {volume} {7}} (\bibinfo {year} {2013})}\BibitemShut {NoStop}%
\bibitem [{\citenamefont {McKee}(2024)}]{mckee2024magnetohydrodynamic}%
  \BibitemOpen
  \bibfield  {author} {\bibinfo {author} {\bibfnamefont {Kyle~I}\ \bibnamefont {McKee}},\ }\bibfield  {title} {\enquote {\bibinfo {title} {Magnetohydrodynamic flow control in {H}ele-{S}haw cells},}\ }\href@noop {} {\bibfield  {journal} {\bibinfo  {journal} {Journal of Fluid Mechanics}\ }\textbf {\bibinfo {volume} {993}},\ \bibinfo {pages} {A11} (\bibinfo {year} {2024})}\BibitemShut {NoStop}%
\bibitem [{Note1()}]{Note1}%
  \BibitemOpen
  \bibinfo {note} {Note that these flows can be superposed with pressure-driven flows to generate a broad new class of flows.}\BibitemShut {Stop}%
\bibitem [{\citenamefont {{H}ele {S}haw}(1898)}]{hele1898flow}%
  \BibitemOpen
  \bibfield  {author} {\bibinfo {author} {\bibfnamefont {H.~S.}\ \bibnamefont {{H}ele {S}haw}},\ }\bibfield  {title} {\enquote {\bibinfo {title} {The flow of water},}\ }\href@noop {} {\bibfield  {journal} {\bibinfo  {journal} {Nature}\ }\textbf {\bibinfo {volume} {58}},\ \bibinfo {pages} {34--36} (\bibinfo {year} {1898})}\BibitemShut {NoStop}%
\bibitem [{\citenamefont {Pozrikidis}(1994)}]{pozrikidis1994motion}%
  \BibitemOpen
  \bibfield  {author} {\bibinfo {author} {\bibfnamefont {C}~\bibnamefont {Pozrikidis}},\ }\bibfield  {title} {\enquote {\bibinfo {title} {The motion of particles in the {H}ele-{S}haw cell},}\ }\href@noop {} {\bibfield  {journal} {\bibinfo  {journal} {Journal of Fluid Mechanics}\ }\textbf {\bibinfo {volume} {261}},\ \bibinfo {pages} {199--222} (\bibinfo {year} {1994})}\BibitemShut {NoStop}%
\bibitem [{\citenamefont {Booth}\ \emph {et~al.}(2025)\citenamefont {Booth}, \citenamefont {Griffiths},\ and\ \citenamefont {Howell}}]{booth2025motion}%
  \BibitemOpen
  \bibfield  {author} {\bibinfo {author} {\bibfnamefont {DJ}~\bibnamefont {Booth}}, \bibinfo {author} {\bibfnamefont {IM}~\bibnamefont {Griffiths}}, \ and\ \bibinfo {author} {\bibfnamefont {PD}~\bibnamefont {Howell}},\ }\bibfield  {title} {\enquote {\bibinfo {title} {The motion of a bubble in a non-uniform {H}ele-{S}haw flow},}\ }\href@noop {} {\bibfield  {journal} {\bibinfo  {journal} {Proceedings of the Royal Society A}\ }\textbf {\bibinfo {volume} {481}},\ \bibinfo {pages} {20240613} (\bibinfo {year} {2025})}\BibitemShut {NoStop}%
\bibitem [{Note2()}]{Note2}%
  \BibitemOpen
  \bibinfo {note} {We consider Hele-Shaw cells with homogeneous electrical properties and cell thickness; cells with heterogeneous properties can support vortical flows\cite {mirzadeh2020vortices,boyko2021microscale,ajdari1995electro,ajdari2000pumping}.}\BibitemShut {Stop}%
\bibitem [{\citenamefont {Van~Dyke}(1982)}]{van1982album}%
  \BibitemOpen
  \bibfield  {author} {\bibinfo {author} {\bibfnamefont {Milton}\ \bibnamefont {Van~Dyke}},\ }\href@noop {} {\emph {\bibinfo {title} {An album of fluid motion}}}\ (\bibinfo  {publisher} {Parabolic Press Stanford},\ \bibinfo {year} {1982})\BibitemShut {NoStop}%
\bibitem [{\citenamefont {McKee}\ and\ \citenamefont {Bush}(2025)}]{mckee2025potential}%
  \BibitemOpen
  \bibfield  {author} {\bibinfo {author} {\bibfnamefont {Kyle~I}\ \bibnamefont {McKee}}\ and\ \bibinfo {author} {\bibfnamefont {John~WM}\ \bibnamefont {Bush}},\ }\bibfield  {title} {\enquote {\bibinfo {title} {Potential flows with electromagnetically induced circulation in a {H}ele-{S}haw cell},}\ }\href@noop {} {\bibfield  {journal} {\bibinfo  {journal} {Physical Review Fluids}\ }\textbf {\bibinfo {volume} {10}},\ \bibinfo {pages} {054103} (\bibinfo {year} {2025})}\BibitemShut {NoStop}%
\bibitem [{Note3()}]{Note3}%
  \BibitemOpen
  \bibinfo {note} {One interesting setting where potential flow with circulation emerges is in rapidly rotating flow satisfying quasi-geostrophic balance. Away from patches of relative vorticity, the flow is a potential flow with circulation; the circulation magnitude around each patch is prescribed by its net vorticity content \cite {pullin1992contour}.}\BibitemShut {Stop}%
\bibitem [{Note4()}]{Note4}%
  \BibitemOpen
  \bibinfo {note} {In practice the upper limit of the $j$-summation is truncated at $N_\protect \mathrm {L}$ terms. In canonical geometries, the solution can be written exactly in terms of the Schottky-Klein prime function\cite {crowdy2020solving}.}\BibitemShut {Stop}%
\bibitem [{Note5()}]{Note5}%
  \BibitemOpen
  \bibinfo {note} {For example, a free stream $U\in \protect \mathbb {C}$ can be incorporated by augmenting $W(z)$ in \protect \eqref {eq:laurent} with the term $\protect \overline {U}z$.}\BibitemShut {Stop}%
\bibitem [{Note6()}]{Note6}%
  \BibitemOpen
  \bibinfo {note} {The semicolon signals vertical stacking.}\BibitemShut {Stop}%
\bibitem [{Note7()}]{Note7}%
  \BibitemOpen
  \bibinfo {note} {Symmetry of $M$ follows from the reciprocity of the Laplace equation, stated as follows, $\DOTSI \intop \ilimits@ _{\partial B_j}\psi _k(\partial B_j)\protect \frac {\partial \psi _j}{\partial n}ds=\DOTSI \intop \ilimits@ _{\partial B_k}\psi _j(\partial B_k)\protect \frac {\partial \psi _k}{\partial n}ds$. Noting that the circulation is unity $\DOTSI \intop \ilimits@ _{\partial B_k}\protect \frac {\partial \psi _k}{\partial n}ds=1$, one finds that $\psi _j(\partial B_k)=\psi _k(\partial B_j)$ or $M_{jk}=M_{kj}$. Positive definiteness follows from Appendix \ref {Powerderiv}.}\BibitemShut {Stop}%
\bibitem [{Note8()}]{Note8}%
  \BibitemOpen
  \bibinfo {note} {$\protect \mathrm {Re}\left \{\protect \overline {W_k'(z)}\right \}=0$ for $z\in \protect \mathbb {R}$ and all $k\in \{1,2,\protect \cdots , N-1\}$}\BibitemShut {NoStop}%
\bibitem [{Note9()}]{Note9}%
  \BibitemOpen
  \bibinfo {note} {In any geometry wherein all obstacles intersect the real axis, while possessing reflectional symmetry about this axis (i.e., $D=\protect \overline {D}$), $\sigma _{\protect \mathrm {max}}$ is necessarily singular along this line. This phenomena is thus by no means restricted to the case of circular conductors.}\BibitemShut {Stop}%
\bibitem [{\citenamefont {Pontryagin}(2018)}]{pontryagin2018mathematical}%
  \BibitemOpen
  \bibfield  {author} {\bibinfo {author} {\bibfnamefont {Lev~Semenovich}\ \bibnamefont {Pontryagin}},\ }\href@noop {} {\emph {\bibinfo {title} {Mathematical theory of optimal processes}}}\ (\bibinfo  {publisher} {Routledge},\ \bibinfo {year} {2018})\BibitemShut {NoStop}%
\bibitem [{\citenamefont {Carroll}(2019)}]{carroll2019spacetime}%
  \BibitemOpen
  \bibfield  {author} {\bibinfo {author} {\bibfnamefont {Sean~M}\ \bibnamefont {Carroll}},\ }\href@noop {} {\emph {\bibinfo {title} {Spacetime and geometry}}}\ (\bibinfo  {publisher} {Cambridge University Press},\ \bibinfo {year} {2019})\BibitemShut {NoStop}%
\bibitem [{Note10()}]{Note10}%
  \BibitemOpen
  \bibinfo {note} {Consider a decomposition of the metric, near a point on the boundary of a conductor, into tangent and normal directions to its surface. Let $n$ denote the normal distance from the boundary and $t$ the tangent distance along it. Since $g^{-1}=\Omega M^{-1} \Omega ^T$, impermeability implies that locally the normal component of the metric scales as $g^{-1}_{nn}\sim n^2/C+ \protect \it {o}\left (n^2\right )$ for some constant $C$. Meanwhile, there is no metric singularity in the tangent direction, $g^{-1}_{tt}\sim C'+ \protect \it {o}\left (1\right )$. Hence, $g_{nn}\sim C/n^2$. Hence the length measured orthogonal to the boundary between $n=\epsilon $ to $n=x$ becomes $\DOTSI \intop \ilimits@ _{\epsilon }^{x}\protect \sqrt {g_{nn}}dn=\DOTSI \intop \ilimits@ _{\epsilon }^{x}\left (1/n\right )dn=\log {\left (x/\epsilon \right )}$. Normal distances thus blowup logarithmically as the boundary is approached.}\BibitemShut {Stop}%
\bibitem [{\citenamefont {Nash}(1954)}]{nash1954c}%
  \BibitemOpen
  \bibfield  {author} {\bibinfo {author} {\bibfnamefont {John}\ \bibnamefont {Nash}},\ }\bibfield  {title} {\enquote {\bibinfo {title} {C1 isometric imbeddings},}\ }\href@noop {} {\bibfield  {journal} {\bibinfo  {journal} {Annals of mathematics}\ }\textbf {\bibinfo {volume} {60}},\ \bibinfo {pages} {383--396} (\bibinfo {year} {1954})}\BibitemShut {NoStop}%
\bibitem [{\citenamefont {Han}\ \emph {et~al.}(2006)\citenamefont {Han}, \citenamefont {Hong},\ and\ \citenamefont {Hong}}]{han2006isometric}%
  \BibitemOpen
  \bibfield  {author} {\bibinfo {author} {\bibfnamefont {Qing}\ \bibnamefont {Han}}, \bibinfo {author} {\bibfnamefont {Jia-Xing}\ \bibnamefont {Hong}}, \ and\ \bibinfo {author} {\bibfnamefont {Jiaxing}\ \bibnamefont {Hong}},\ }\href@noop {} {\emph {\bibinfo {title} {Isometric embedding of Riemannian manifolds in Euclidean spaces}}},\ Vol.~\bibinfo {volume} {13}\ (\bibinfo  {publisher} {American Mathematical Soc.},\ \bibinfo {year} {2006})\BibitemShut {NoStop}%
\bibitem [{\citenamefont {Chern}\ \emph {et~al.}(2018)\citenamefont {Chern}, \citenamefont {Kn{\"o}ppel}, \citenamefont {Pinkall},\ and\ \citenamefont {Schr{\"o}der}}]{chern2018shape}%
  \BibitemOpen
  \bibfield  {author} {\bibinfo {author} {\bibfnamefont {Albert}\ \bibnamefont {Chern}}, \bibinfo {author} {\bibfnamefont {Felix}\ \bibnamefont {Kn{\"o}ppel}}, \bibinfo {author} {\bibfnamefont {Ulrich}\ \bibnamefont {Pinkall}}, \ and\ \bibinfo {author} {\bibfnamefont {Peter}\ \bibnamefont {Schr{\"o}der}},\ }\bibfield  {title} {\enquote {\bibinfo {title} {Shape from metric.}}\ }\href@noop {} {\bibfield  {journal} {\bibinfo  {journal} {ACM Trans. Graph.}\ }\textbf {\bibinfo {volume} {37}},\ \bibinfo {pages} {63} (\bibinfo {year} {2018})}\BibitemShut {NoStop}%
\bibitem [{Note11()}]{Note11}%
  \BibitemOpen
  \bibinfo {note} {Sampling $M^{\protect \frac {1}{2}}\protect \bm {\Gamma }$ rather than $\protect \bm {\Gamma }$ from a mean-zero Gaussian distribution ensures that equal-power ($P=\protect \bm {\Gamma }^T M \protect \bm {\Gamma }$) configurations are explored equally. When the distribution has variance $\sigma ^2$, the electrical power follows a Chi-squared distribution with $\protect \mathbb {E}(P)=\left (N-1\right )\sigma ^2$.}\BibitemShut {Stop}%
\bibitem [{Note12()}]{Note12}%
  \BibitemOpen
  \bibinfo {note} {Our numerical method utilized the Hamiltonian form of the Euler-Lagrange equations \protect \eqref {eq:geodesic} wherein $H=\left (\protect \bm {p}^Tg^{-1}\protect \bm {p}\right )/4$ and $\protect \bm {p}=2g\protect \dot {\protect \bm {x}}$. In this form, we verify convergence of our integrations by monitoring the Hamiltonian conservation along trajectories; for geodesics included in the paper, Hamiltonian fluctuations are less that 1\% along trajectories.}\BibitemShut {Stop}%
\bibitem [{Note13()}]{Note13}%
  \BibitemOpen
  \bibinfo {note} {In the 2D magnetic system, $N=3$ conductors really only corresponds to two degrees of freedom because of the current conservation constraint.}\BibitemShut {Stop}%
\bibitem [{\citenamefont {Qin}\ and\ \citenamefont {Arratia}(2022)}]{qin2022confinement}%
  \BibitemOpen
  \bibfield  {author} {\bibinfo {author} {\bibfnamefont {Boyang}\ \bibnamefont {Qin}}\ and\ \bibinfo {author} {\bibfnamefont {Paulo~E}\ \bibnamefont {Arratia}},\ }\bibfield  {title} {\enquote {\bibinfo {title} {Confinement, chaotic transport, and trapping of active swimmers in time-periodic flows},}\ }\href@noop {} {\bibfield  {journal} {\bibinfo  {journal} {Science Advances}\ }\textbf {\bibinfo {volume} {8}},\ \bibinfo {pages} {eadd6196} (\bibinfo {year} {2022})}\BibitemShut {NoStop}%
\bibitem [{\citenamefont {Uchytil}\ \emph {et~al.}(2025)\citenamefont {Uchytil}, \citenamefont {Korda},\ and\ \citenamefont {Zem{\'a}nek}}]{uchytil2025data}%
  \BibitemOpen
  \bibfield  {author} {\bibinfo {author} {\bibfnamefont {Adam}\ \bibnamefont {Uchytil}}, \bibinfo {author} {\bibfnamefont {Milan}\ \bibnamefont {Korda}}, \ and\ \bibinfo {author} {\bibfnamefont {Ji{\v{r}}{\'\i}}\ \bibnamefont {Zem{\'a}nek}},\ }\bibfield  {title} {\enquote {\bibinfo {title} {Data-driven control of a magnetohydrodynamic flow},}\ }\href@noop {} {\bibfield  {journal} {\bibinfo  {journal} {arXiv preprint arXiv:2507.12479}\ } (\bibinfo {year} {2025})}\BibitemShut {NoStop}%
\bibitem [{\citenamefont {Karnakov}\ \emph {et~al.}(2025)\citenamefont {Karnakov}, \citenamefont {Amoudruz},\ and\ \citenamefont {Koumoutsakos}}]{karnakov2025optimal}%
  \BibitemOpen
  \bibfield  {author} {\bibinfo {author} {\bibfnamefont {Petr}\ \bibnamefont {Karnakov}}, \bibinfo {author} {\bibfnamefont {Lucas}\ \bibnamefont {Amoudruz}}, \ and\ \bibinfo {author} {\bibfnamefont {Petros}\ \bibnamefont {Koumoutsakos}},\ }\bibfield  {title} {\enquote {\bibinfo {title} {Optimal navigation in microfluidics via the optimization of a discrete loss},}\ }\href@noop {} {\bibfield  {journal} {\bibinfo  {journal} {Physical Review Letters}\ }\textbf {\bibinfo {volume} {134}},\ \bibinfo {pages} {044001} (\bibinfo {year} {2025})}\BibitemShut {NoStop}%
\bibitem [{\citenamefont {Mirzadeh}\ \emph {et~al.}(2020)\citenamefont {Mirzadeh}, \citenamefont {Zhou}, \citenamefont {Amooie}, \citenamefont {Fraggedakis}, \citenamefont {Ferguson},\ and\ \citenamefont {Bazant}}]{mirzadeh2020vortices}%
  \BibitemOpen
  \bibfield  {author} {\bibinfo {author} {\bibfnamefont {Mohammad}\ \bibnamefont {Mirzadeh}}, \bibinfo {author} {\bibfnamefont {Tingtao}\ \bibnamefont {Zhou}}, \bibinfo {author} {\bibfnamefont {Mohammad~Amin}\ \bibnamefont {Amooie}}, \bibinfo {author} {\bibfnamefont {Dimitrios}\ \bibnamefont {Fraggedakis}}, \bibinfo {author} {\bibfnamefont {Todd~R}\ \bibnamefont {Ferguson}}, \ and\ \bibinfo {author} {\bibfnamefont {Martin~Z}\ \bibnamefont {Bazant}},\ }\bibfield  {title} {\enquote {\bibinfo {title} {Vortices of electro-osmotic flow in heterogeneous porous media},}\ }\href@noop {} {\bibfield  {journal} {\bibinfo  {journal} {Physical Review Fluids}\ }\textbf {\bibinfo {volume} {5}},\ \bibinfo {pages} {103701} (\bibinfo {year} {2020})}\BibitemShut {NoStop}%
\bibitem [{\citenamefont {Boyko}\ \emph {et~al.}(2021)\citenamefont {Boyko}, \citenamefont {Bacheva}, \citenamefont {Eigenbrod}, \citenamefont {Paratore}, \citenamefont {Gat}, \citenamefont {Hardt},\ and\ \citenamefont {Bercovici}}]{boyko2021microscale}%
  \BibitemOpen
  \bibfield  {author} {\bibinfo {author} {\bibfnamefont {Evgeniy}\ \bibnamefont {Boyko}}, \bibinfo {author} {\bibfnamefont {Vesna}\ \bibnamefont {Bacheva}}, \bibinfo {author} {\bibfnamefont {Michael}\ \bibnamefont {Eigenbrod}}, \bibinfo {author} {\bibfnamefont {Federico}\ \bibnamefont {Paratore}}, \bibinfo {author} {\bibfnamefont {Amir~D}\ \bibnamefont {Gat}}, \bibinfo {author} {\bibfnamefont {Steffen}\ \bibnamefont {Hardt}}, \ and\ \bibinfo {author} {\bibfnamefont {Moran}\ \bibnamefont {Bercovici}},\ }\bibfield  {title} {\enquote {\bibinfo {title} {Microscale hydrodynamic cloaking and shielding via electro-osmosis},}\ }\href@noop {} {\bibfield  {journal} {\bibinfo  {journal} {Physical Review Letters}\ }\textbf {\bibinfo {volume} {126}},\ \bibinfo {pages} {184502} (\bibinfo {year} {2021})}\BibitemShut {NoStop}%
\bibitem [{\citenamefont {Ajdari}(1995)}]{ajdari1995electro}%
  \BibitemOpen
  \bibfield  {author} {\bibinfo {author} {\bibfnamefont {Armand}\ \bibnamefont {Ajdari}},\ }\bibfield  {title} {\enquote {\bibinfo {title} {Electro-osmosis on inhomogeneously charged surfaces},}\ }\href@noop {} {\bibfield  {journal} {\bibinfo  {journal} {Physical Review Letters}\ }\textbf {\bibinfo {volume} {75}},\ \bibinfo {pages} {755} (\bibinfo {year} {1995})}\BibitemShut {NoStop}%
\bibitem [{\citenamefont {Ajdari}(2000)}]{ajdari2000pumping}%
  \BibitemOpen
  \bibfield  {author} {\bibinfo {author} {\bibfnamefont {Armand}\ \bibnamefont {Ajdari}},\ }\bibfield  {title} {\enquote {\bibinfo {title} {Pumping liquids using asymmetric electrode arrays},}\ }\href@noop {} {\bibfield  {journal} {\bibinfo  {journal} {Physical review E}\ }\textbf {\bibinfo {volume} {61}},\ \bibinfo {pages} {R45} (\bibinfo {year} {2000})}\BibitemShut {NoStop}%
\bibitem [{\citenamefont {Pullin}(1992)}]{pullin1992contour}%
  \BibitemOpen
  \bibfield  {author} {\bibinfo {author} {\bibfnamefont {DI}~\bibnamefont {Pullin}},\ }\bibfield  {title} {\enquote {\bibinfo {title} {Contour dynamics methods},}\ }\href@noop {} {\bibfield  {journal} {\bibinfo  {journal} {Annual review of fluid mechanics}\ }\textbf {\bibinfo {volume} {24}},\ \bibinfo {pages} {89--115} (\bibinfo {year} {1992})}\BibitemShut {NoStop}%
\bibitem [{\citenamefont {Crowdy}(2020)}]{crowdy2020solving}%
  \BibitemOpen
  \bibfield  {author} {\bibinfo {author} {\bibfnamefont {D.}~\bibnamefont {Crowdy}},\ }\href@noop {} {\emph {\bibinfo {title} {Solving problems in multiply connected domains}}}\ (\bibinfo  {publisher} {SIAM},\ \bibinfo {year} {2020})\BibitemShut {NoStop}%
\bibitem [{\citenamefont {Kim}\ and\ \citenamefont {Karrila}(2013)}]{kim2013microhydrodynamics}%
  \BibitemOpen
  \bibfield  {author} {\bibinfo {author} {\bibfnamefont {Sangtae}\ \bibnamefont {Kim}}\ and\ \bibinfo {author} {\bibfnamefont {Seppo~J}\ \bibnamefont {Karrila}},\ }\href@noop {} {\emph {\bibinfo {title} {Microhydrodynamics: principles and selected applications}}}\ (\bibinfo  {publisher} {Butterworth-Heinemann},\ \bibinfo {year} {2013})\BibitemShut {NoStop}%
\end{thebibliography}%
\newpage
\appendix
\section{Electrical Power Expression} \label{Powerderiv}
The electrical dissipation in the fluid domain is given by,
\begin{equation}  P=\int_{D}\boldsymbol{J}\boldsymbol{\cdot}\boldsymbol{E}=\sigma\int_{D}\nabla V(\boldsymbol{x})\cdot \nabla V(\boldsymbol{x}) dV,
\end{equation}
which is clearly positive definite. Noting that $V$ is harmonic, $\nabla^2V=0$, one finds via the divergence theorem that
\begin{equation}
P=h\sum_{k=1}^{N}\sigma V_k \int_{\partial B_k}\sigma \nabla V(\boldsymbol{x})\cdot \boldsymbol{n}dA=h\sum_{k=1}^{N} V_k I_k,
\end{equation}
where $\boldsymbol{n}$ is the normal direction pointing out of the fluid so that $I_k$ is the electrical current exiting the $j^{\mathrm{th}}$ body into the fluid. Noting that the depth-averaged circulation is given by $\Gamma_k = - h B_0 I_k/\left(12\mu\right)$ from \eqref{eq:circulaitonfo}, the electrical power can be written as 
\begin{equation}
P=-\frac{12\mu}{B_0}\sum_{k=1}^{N-1} V_k \Gamma_{k}.
\end{equation}
Without loss of generality, the $N^{\mathrm{th}}$ term of the summation has been ignored since as we fix $V_N=0$ in our boundary value problem. The voltage $V_k$ can be written in terms of the known flow solution,
\begin{equation}
    V_k=\frac{12\mu}{h^2\sigma B_0}\sum_{j=1}^{N-1} \Gamma_j \mathrm{Im}\left\{ W_j(\partial B_k)\right\},
\end{equation}
so that the power can be written as,
\begin{equation}
P=-\frac{144\mu^2}{h^2\sigma B_0^2}\sum_{k=1}^{N-1}\sum_{j=1}^{N-1} \Gamma_k\Gamma_j \mathrm{Im}\left\{ W_j(\partial B_k)\right\}.
\end{equation}

\section{Power-Minimizing Control via Lagrange Multiplier Approach}\label{LagrangeDeriv}
To find the power-minimizing solution, we use a Lagrange multiplier approach and examine the following function,
\begin{equation}  f(\boldsymbol{x},\boldsymbol{\lambda})=\boldsymbol{\Gamma}^T M \boldsymbol{\Gamma}+\boldsymbol{\lambda}^T \left(\Omega\boldsymbol{\Gamma}-\dot{\boldsymbol{x}}\right),
\end{equation}
where $\boldsymbol{\lambda}$ represents our Lagrange multiplier vector.

Differentiating with respect to $\boldsymbol{\Gamma}$ and setting to zero, we find,
\begin{equation}\label{eq:dgam}
2M\boldsymbol{\Gamma}=- \Omega^T\boldsymbol{\lambda}.
\end{equation}
Plugging into the constraint equation, we find that,
\begin{equation}
\boldsymbol{\lambda}=-2\left(\Omega M^{-1}\Omega^T\right)^{-1}\dot{\boldsymbol{x}},
\end{equation}
which together with (\ref{eq:dgam}), yields the power-minimizing solution,
\begin{equation}
    \boldsymbol{\Gamma}=M^{-1}\Omega^T\left(\Omega M^{-1} \Omega^T\right)^{-1}\dot{\boldsymbol{x}}.
\end{equation}

\section{Derivation of Geodesic Equation}\label{geodesicderivations}
Considering the Lagrangian $\mathcal{L}=g_{kl}\dot{x}_k \dot{x}_l$, the Euler-Lagrange equations $d\left(\partial \mathcal{L}/\partial \dot{x}_p\right)/dt-\partial \mathcal{L}/\partial x_p=0$ become,
\begin{eqnarray}
    0=\frac{d}{dt}\left(2g_{k p }\dot{x}_k\right)-\frac{\partial g_{k l}}{\partial x_p}\dot{x}_k\dot{x}_l\\
    0=2\frac{\partial g_{k p }}{\partial x_q}\dot{x}_q\dot{x}_k+2g_{k p }\ddot{x}_k-\frac{\partial g_{k l}}{\partial x_p}\dot{x}_k\dot{x}_l.
\end{eqnarray}
By noting that for any real square matrix $B$, $\boldsymbol{x}^T B \boldsymbol{x}=\boldsymbol{x}^T \left(B+B^T\right) \boldsymbol{x}/2$, we find that the latter equation reduces to,
\begin{equation}
0=\frac{\partial g_{q p }}{\partial x_k}\dot{x}_q\dot{x}_k+\frac{\partial g_{k p }}{\partial x_q}\dot{x}_q\dot{x}_k+2g_{k p }\ddot{x}_k-\frac{\partial g_{k l}}{\partial x_p}\dot{x}_k\dot{x}_l.
\end{equation}
Multiplying by the inverse of the metric and rearranging dummy indices gives,
\begin{equation}
\ddot{x}_m=-\frac{1}{2}g_{mp}^{-1}\left(\frac{\partial g_{q p }}{\partial x_k}+\frac{\partial g_{k p }}{\partial x_q} -\frac{\partial g_{k q}}{\partial x_p}\right)\dot{x}_k\dot{x}_q,
\end{equation}
from which \eqref{eq:geodesic} follows.

\section{Proof of Time-Optimality}\label{timeoptimality}
We herein deduce the form of the trajectory that minimizes the transport time between two points, subject to maximum power constraint of the form $P(\tau) \leq P_{\mathrm{max}}$ for all time. The proof is laid out in two parts. We first deduce that at each moment, it is sufficient to consider the power-minimizing control \eqref{eq:powermincontrolfun} when seeking a time-optimal solution. Then, restricting our attention to such controls, we prove that the geodesic trajectory is time-optimal and that it should be traversed with constant power $P_{\mathrm{max}}$.

First, we note that when controlling a trajectory $\boldsymbol{x}(t)$, the circulation only takes the form \eqref{eq:powermincontrolfun} if we choose to minimize instantaneous power. It is not immediately clear that this choice should also correspond to time-optimality, though we will now show this to be the case. Suppose that there were a time-optimal control, corresponding to the trajectory $\boldsymbol{x}(t)$, that utilizes a control different to \eqref{eq:powermincontrolfun}. Let $P_{\mathrm{o}}(t)$ be the power corresponding to that control, which is achieved over the interval $t\in[0,T_{\mathrm{end}}]$.
Then, the power-optimal control for that trajectory has the property that $\dot{\boldsymbol{x}}^T g \dot{\boldsymbol{x}} \leq P_\mathrm{o}(t)$ for all time. Moreover, consider a time re-parametrization of the  trajectory $\boldsymbol{x}(t)$, given by $\tilde{\boldsymbol{x}}(t)=\boldsymbol{x}(f(t))$. The power required by the power-optimal control to achieve this trajectory is then given by $\left(\dot{f}(t)\right)^2\dot{\boldsymbol{x}}(f(t))^T g \dot{\boldsymbol{x}}(f(t))$ on the interval $t\in [0,T_f]$ with $T_{f}$ defined by $f(T_f)=T_{\mathrm{end}}$. Now we let $f(t)$ by defined by the following initial-value-problem,
\begin{eqnarray}
\dot{f}(t)&=&+\sqrt{\frac{P_{\mathrm{o}}(f(t))}{\dot{\boldsymbol{x}}(f(t))^T  g  \dot{\boldsymbol{x}}(f(t))}},\label{eq:fode}\\
f(0)&=&0.
\end{eqnarray}
Since for all times, $P_{\mathrm{o}}(t)\geq \dot{\boldsymbol{x}}(f(t))^Tg\dot{\boldsymbol{x}}(f(t))$, it follows that $T_{\mathrm{f}}\leq T_{\mathrm{end}}$. By construction, the power associated with this trajectory is given exactly by $P_o(t)$. We have thus shown that given a trajectory $\boldsymbol{x}(t)$ with a generic control and associated power $P_o(t)$, it is always possible to construct a trajectory which uses the power optimal control \eqref{eq:powermincontrolfun} and for the same power expenditure traverses the curve defined by $\boldsymbol{x}(t)$ in lesser or equal time. Thus while seeking the time-optimal control we may restrict our attention to controls that, at each instant, employ the power-optimal control \eqref{eq:powermincontrolfun}. 

Restricting our attention to such controls we now seek the trajectory which, subject to the power constraint $P\leq P_{\mathrm{max}}$, achieves the fastest transport from $\boldsymbol{x}_{\mathrm{A}}$ to $\boldsymbol{x}_{\mathrm{B}}$. Suppose that the time optimal trajectory is given by $\boldsymbol{x}(\tau)$ over the time interval $[0,T]$ where $\boldsymbol{x}(0)=\boldsymbol{x}_A$ and $\boldsymbol{x}(T)=\boldsymbol{x}_B$. Suppose now that $\boldsymbol{x}(\tau)$ is \emph{not} derived from the geodesic equation laid out in \eqref{eq:geodesic}, and we shall proceed toward a contradiction. Consider the energy functional,
\begin{eqnarray}
    \epsilon [\boldsymbol{x}]&=&\int_{0}^{T}\dot{\boldsymbol{x}}(\tau)^T g(\boldsymbol{x}(\tau)) \dot{\boldsymbol{x}}(\tau)d\tau\\
    \boldsymbol{x}(0)&=&\boldsymbol{x}_{A}\\
    \boldsymbol{x}(T)&=&\boldsymbol{x}_{B},
\end{eqnarray}
whose energy-optimal ($\epsilon$-minimizing) solution is given by the Euler-Lagrange equations. Let $\boldsymbol{x}_\mathrm{geod}$ be the corresponding geodesic trajectory. Since this trajectory optimizes the energy among all trajectories, we can be sure that $\epsilon[\boldsymbol{x}_{\mathrm{geod}}]\leq \epsilon[\boldsymbol{x}]$. By the constraint of our problem statement, we also have $\epsilon[\boldsymbol{x}]\leq P_{\mathrm{max}}T$. Noting that geodesics correspond to constant-power trajectories (since the Hamiltonian and Lagrangian are equal and time-independent), we have that $\epsilon[\boldsymbol{x}_{\mathrm{geod}}]=P_{\mathrm{geod}}T$, where $P_{\mathrm{geod}}$ is the constant geodesic power. Therefore the inequality $\epsilon[\boldsymbol{x}_{\mathrm{geod}}]\leq \epsilon[\boldsymbol{x}]$ implies the weakened inequality $P_{\mathrm{geod}}\leq P_{\mathrm{max}}$. Thus, a geodesic path can always be found that both (1) satisfies the power constraint and (2) connects two points in equal or lesser time as compared to $\boldsymbol{x}(\tau)$. Moreover, the fastest geodesic is found by taking $P_{\mathrm{geod}}= P_{\mathrm{max}}$.

We thus deduce that \emph{the time-optimal trajectory under the constraint that $P(t)\leq P_{\mathrm{max}}$ is the geodesic curve traversed with $P(t)= P_{\mathrm{max}}$.} To achieve this particular trajectory in practice, one can initialize velocities in the shooting method for finding geodesics so that the power condition is initially satisfied, $\dot{\boldsymbol{x}}(0)^T g(\boldsymbol{x}_{\mathrm{A}})\dot{\boldsymbol{x}}(0)=P_{\mathrm{max}}$, which will then be preserved automatically by the conservative dynamics.

\section{Three-dimensional Stokes Flow -- Rotating Disks in a Cube}\label{stokesrotat}
Consider Stokes flow inside a cube $C\subset \mathbb{R}^3$ with rotating disks mounted on $N_d$ of its walls, with controllable rotation rates $\boldsymbol{\omega}=(\omega_1,\cdots,\omega_{N_d})^T$. Let $\partial C_k$ denote the surface of the $k^{\mathrm{th}}$ rotating disk for $k\in\{1,\cdots, N_d\}$. The fluid flow can then be written as $\boldsymbol{u}(\boldsymbol{x})=\sum_{k=1}^{N_d} \omega_k \boldsymbol{u}_k(\boldsymbol{x})$, where $\boldsymbol{u}_k(\boldsymbol{x})$ is the flow corresponding to $\omega_k=1$ and $\omega_{j\neq k}=0$. We then define $\Omega\in \mathbb{R}^{3\times N_d}$, by $\Omega=(\boldsymbol{u}_1, \cdots, \boldsymbol{u}_{N_d})$. The control problem is then given by \eqref{eq:controleq}, where now $\Omega\in \mathbb{R}^{3\times N_d}$ and $\boldsymbol{x}\in C$.

The power required to generate a flow $\boldsymbol{u}$ is given by $\int_{\partial C}\boldsymbol{u}\boldsymbol{\cdot} \boldsymbol{\sigma} \boldsymbol{\cdot} \boldsymbol{n} dS$, with $\boldsymbol{\sigma} =\mu\left(\nabla \boldsymbol{u}+\nabla \boldsymbol{u}^T\right)$, where pressure has dropped out because boundaries only \emph{rotate}, from which it follows that $\boldsymbol{u}\boldsymbol{\cdot}\boldsymbol{n}=0$ on all surfaces. Substituting our expression for the velocity field, we find that,
\begin{eqnarray}
P=\sum_{q=1}^{N_d}\sum_{r=1}^{N_d}\mu\omega_q\omega_rM_{qr}\equiv \mu \boldsymbol{\omega}^T M \boldsymbol{\omega}, \label{eq:stokespower}\\
M_{qr}=\int_{\partial C_q}\boldsymbol{u}_q\boldsymbol{\cdot}\left(\nabla \boldsymbol{u}_r+\nabla \boldsymbol{u}_r^T\right)\boldsymbol{\cdot}\boldsymbol{n}dS,\label{eq:StokesM}
\end{eqnarray}
where $M$ is positive definite and symmetric. Positive definiteness follows from the fact \eqref{eq:stokespower} is equal to the integral of a positive definite quantity over $C$. Symmetry follows from the reciprocal theorem \cite{kim2013microhydrodynamics}. With these definitions of $\Omega$ and $M$, equations \eqref{eq:powermincontrolfun} and \eqref{eq:geodesic} hold with a metric still defined by $g=\left(\Omega M^{-1} \Omega^T\right)^{-1}$. Particles forced by random boundary disk rotations then diffuse according to the anisotropic diffusion tensor $g^{-1}=\Omega M^{-1} \Omega^T$.
\end{document}